\documentstyle[emulateapj,apjfonts,psfig,natbib,amsmath]{article}

\begin{document}

\def\cit{1}
\def\vla{2}
\def\atca{3}

\title{Constraints on Off-Axis GRB Jets in Type Ibc Supernovae From Late-Time Radio Observations}

\author{A. M. Soderberg \altaffilmark{\cit},
D. A. Frail\altaffilmark{\vla},
\&\ M. H. Wieringa\altaffilmark{\atca}
}

\altaffiltext{\cit}{Division of Physics, Mathematics and Astronomy,
        105-24, California Institute of Technology, Pasadena, CA
        91125}
\altaffiltext{\vla}{National Radio Astronomy Observatory, Socorro,
        NM 87801}
\altaffiltext{\atca}{Australia Telescope National Facility, CSIRO,
P.O. Box 76, Epping NSW 1710, Australia}

\begin{abstract}
   It has been suggested that the peculiar properties of the luminous
   Type Ic supernova SN\,1998bw and its low-energy gamma-ray
   burst GRB\,980425 may be understood if they originated in a
   standard gamma-ray burst explosion viewed far from the axis of the
   relativistic jet.  In this scenario, strong radio emission is
   predicted from the jet on a timescale 1 to 10 years after the
   explosion as it decelerates and spreads into our line of sight.  To
   test this hypothesis we have carried out late-time radio
   observations of SN\,1998bw at $t=5.6$ years, yielding
   upper limits which are consistent with the continued fading of the
   supernova. We find these limits to be consistent with an off-axis
   jet only if the progenitor mass loss rate is $\dot{M}\lesssim 4
   \times 10^{-7}$ M$_\odot$ yr$^{-1}$ (for a wind velocity $v_w=1000$
   km s$^{-1}$) or the fraction of the shock energy in magnetic fields
   is $\epsilon_B \lesssim 10^{-3}$.  These values are low relative
   to those inferred for cosmological GRBs.  We combine the SN\,1998bw
   measurements with existing observations for a sample of 15 local
   Type Ibc supernovae to estimate that at most 6\% produce
   collimated, relativistic outflows.
\end{abstract}


\section{Introduction}

GRB\,980425 was the first gamma-ray burst (GRB) to be identified with
a supernova (SN), SN\,1998bw \citep{gvv+98,paa+00}.  However, this
association was not universally accepted since the spatial
and temporal coincidence of these events could only be approximated to
$\pm 8^\prime$ and $\pm$2 d, respectively. Recently, the detection
of spectroscopic features in the light curve of GRB\,030329, similar
to those seen in SN\,1998bw \citep{smg+03,hsm+03}, has strengthened
the hypothesis of the SN\,1998bw/GRB\,980425 association.
These results have given a powerful impetus to the ``collapsar'' model
\citep{mw99,zwm03} in which a Wolf-Rayet progenitor undergoes
core collapse, producing a rapidly rotating black hole surrounded by
an accretion disk which injects energy into the system and thus acts
as a ``central engine''. The energy extracted from this system supports a
quasi-spherical Type Ibc SN explosion and drives 
collimated jets through the stellar rotation axis which produce
the prompt gamma-ray and afterglow emission (see review by
\citealt{zm03}).

Despite this progress, some unresolved issues remain.  Foremost among
these is understanding the connection between local SNe Ibc and
cosmological GRBs.  Estimates of the fraction of Type Ibc SNe that
produce a GRB range from $10^{-5}$ \citep{psf03}, to 0.5\%
\citep{fks+01}, and even approaching 100\% \citep{ldg03}.  This
uncertainty in the relative rates of Ibc SNe and GRBs is due to
different assumptions about the geometry and energetics of GRBs.

As the only local SN observed in association with a GRB, SN\,1998bw is
the key to our understanding.  At a distance of 38 Mpc, this Type Ic
SN was unusually luminous at optical and radio wavelengths
\citep{kfw+98,gvv+98}.  The broad absorption lines seen in early
spectra of SN\,1998bw implied expansion velocities of $\ge
30,000\rm~km s^{-1}$ and (isotropic) kinetic energies of $\sim 3
\times 10^{52}$ erg \citep{imn+98,wes99,tmm01}. Likewise, the bright,
early-peaked radio emission implied a significant amount of energy
($\sim 10^{50}$ erg) coupled to mildly ($\Gamma\approx 2$)
relativistic ejecta \citep{kfw+98}. Detailed modeling by \citet{lc99}
confirmed this result and also showed the need for a second energy
injection episode, indicating the presence of a central engine,
similar to the model inferred for GRBs.

However, in contrast to the large energies inferred from the optical and radio
emission from SN\,1998bw, GRB\,980425 was a sub-energetic
gamma-ray burst. The prompt emission had an (isotropic) energy release of
only $8\times 10^{47}$ erg \citep{paa+00} -- 4 to 6 orders of magnitude below
typical GRBs \citep{fks+01}.

One popular explanation posits that SN\,1998bw/GRB980425 was a typical
GRB viewed away from the jet axis
\citep{nak98,cen98,el99,wes99,sal01,yyn03}.  
This hypothesis can been described by two different scenarios
based on the ratio of the off-axis
angle, $\theta_{\rm oa}$, to the opening angle of the GRB jet, $\theta_j$.
In Case 1, $\theta_{\rm oa} \sim 3\theta_j$ so the jet emission
is detected at early time.  Here, the $\gamma$-rays
originate from the edge of the relativistic jet,
causing the inferred (isotropic) energy to be suppressed \citep{gpk+02}.  
In Case 2, $\theta_{\rm oa}$
is large ($\gg 10$ $\theta_j$) so the jet emission is only detectable
at late time ($t\sim 1-10$ years) when the jet has reached spherical symmetry.
In this case, the prompt $\gamma$-ray emission
may be due to Compton scattering of photons into our
line-of-sight \citep{wax03}.  

\citet{gpk+02} have investigated Case 1, finding the optical and
$\gamma$-ray emission to be consistent with an off-axis angle of
$\theta_{\rm oa}\approx 4^o$ (for $\Gamma \approx 100$).  This scenario,
however, may have difficulty explaining the X-ray and radio evolution.
\citet{wax03} has recently investigated Case 2, predicting a late-time
rise in the observed luminosity which is most easily detectable at
radio frequencies \citep{pac01,tot03,gl03}.  

The possibility of Case 2 has given rise to unification models which 
imply some fraction of local Ibc SNe can also be
described as GRBs viewed off-axis \citep{wax03,ldg03}.  While the
off-axis jet model provides a convenient framework in which to unite
the GRB and SN phenomena, confirmation of the model requires
observational evidence for a GRB jet within a local SN.  In this paper
we carry out late-time radio observations of SN\,1998bw in an
effort to detect the putative off-axis jet from GRB\,980425.  We
combine this measurement with existing observations for a sample of
nearby Type Ibc SNe to place constraints on the parameters of
the off-axis jet and the fraction of local Ibc SNe which could
be associated with relativistic jets similar to those seen in cosmological
GRBs.

\section{Observations}
\label{sec:obs}

We observed SN\,1998bw with the Australia Telescope Compact Array
(ATCA)\footnotemark\footnotetext{The Australia Telescope is funded by
the Commonwealth of Australia for operation as a National Facility
managed by CSIRO.}  beginning at 2003 December 4 21:00 UT and ending
on 2003 December 5 8:30 UT. Dual frequency observations were made at
1384 and 2368 MHz, using a bandwidth of 128 MHz and two orthogonal
linear polarizations for each frequency band. The final synthesized
beams were 8$\times 14^{\prime\prime}$ and 5$\times 8^{\prime\prime}$,
at 1384 and 2368 MHz, respectively.

\begin{figure}
\plotone{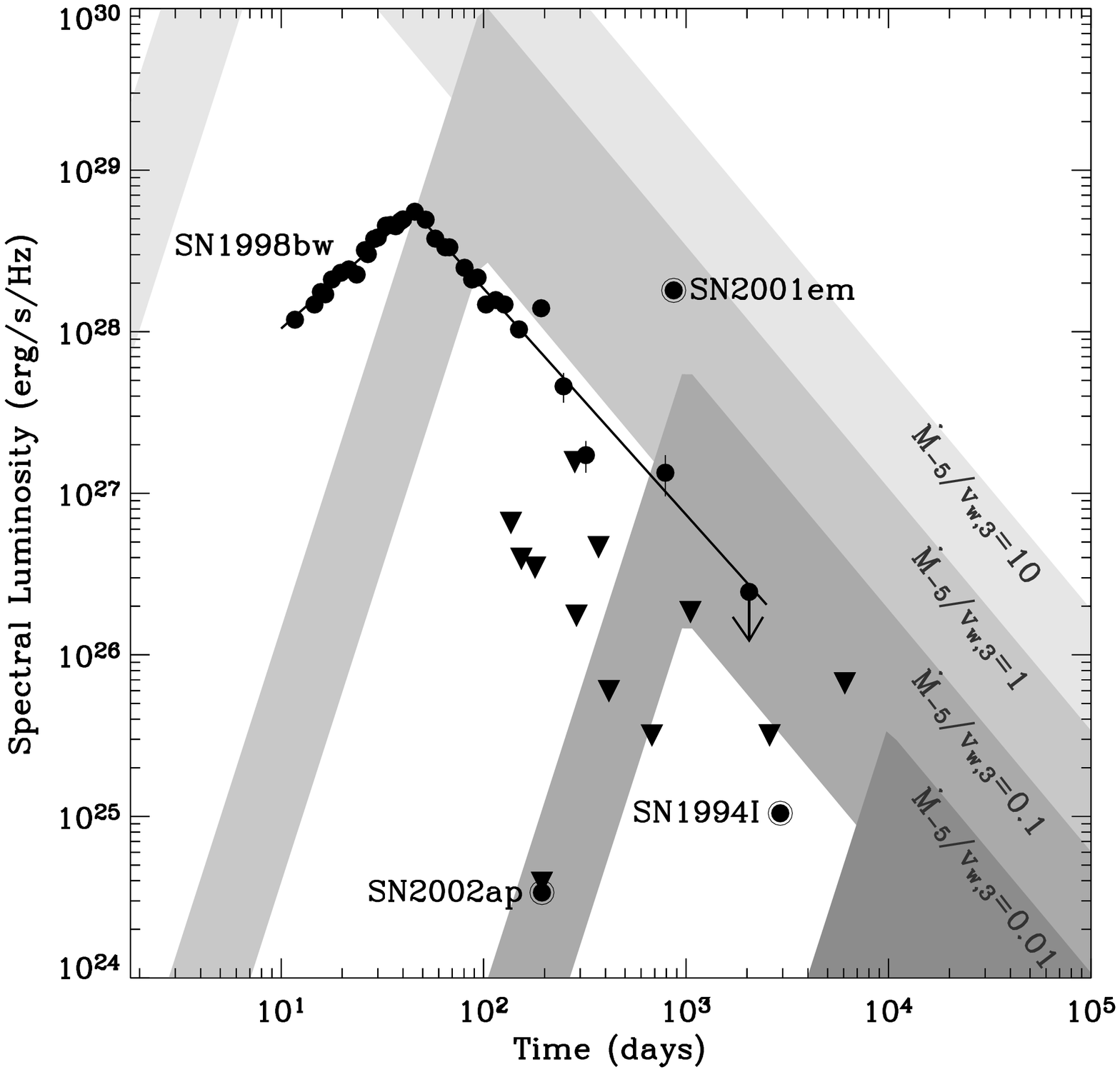}
\caption{The radio luminosity evolution of SN 1998bw is
shown with late-time detection limits for 12 nearby Ibc supernovae
(triangles) and three late-time detections for SN\,
1994I, SN\,2002ap and SN\,2001em.  
The four shaded regions represent an off-axis GRB jet for
$A_*=\dot{M}_{-5}/v_{w,3}=(0.01, 0.1, 1, 10)$ and assuming a range of
values, $\epsilon_B=0.002-0.25$. 
We have assumed typical values of $E_{51}=1$ and
$\epsilon_e=0.1$.  With the exception of SN\,2001em, all of these late-time observations are significantly
fainter than the $\dot{M}_{-5}/v_{w,3}=1$ off-axis jet prediction and
nine of the SNe constrain $\dot{M}_{-5}/v_{w,3} \le 0.1$.}
\label{fig:lum_limits}
\end{figure}

No emission is detected toward SN\,1998bw at either frequency but
there is a weak source (209 $\mu$Jy) detected at 2368 MHz
$1.7^{\prime\prime}$ west and $3.1^{\prime\prime}$ north of the SN.
This source is too bright to be due to thermal emission from the
complex of HII regions in this area. It is possible that this is the
radio counterpart of the ultraluminous X-ray transient (S1b) recently
identified by Kouveliotou et al.~(2004)\nocite{kwp+04} but it would be
60-times more luminous than the only other known detection
\citep{kcp+03}. The peak flux density at the position of the SN was
$62\pm 40$ and $149\pm 50$ $\mu$Jy for 1384 and 2368 MHz,
respectively.  These new upper limits, taken 2049.19 d after
GRB\,980425 exploded ($t\simeq$5.6 yrs), lie just below the
extrapolation of a power-law decay from 50 to 790 days
(Figure~\ref{fig:lum_limits}).

\section{SN\,1998bw: Constraining the Off-Axis Jet Model}
\label{sec:sn98bw}

\citet{wax03} has made predictions for the evolution of late-time
radio emission from a typical GRB jet viewed far from
the jet axis.  As the off-axis jet sweeps up material and decelerates,
it eventually undergoes a dynamical transition to sub-relativistic
expansion \citep{fwk00} and spreads sideways into our line-of-sight.
This occurs on a timescale $t_{\rm oa} \approx 
E_{51}/(\dot{M}_{-5}/v_{w,3})$ years.  Here $E_{51}$ is the energy of
the jet (normalized to $10^{51}$ erg), $v_w$ is the velocity of the
stellar wind, and $\dot{M}_{-5}/v_{w,3}$ is the mass loss rate of the
progenitor star normalized as $(\dot{M}/10^{-5}~M_{\odot}~{\rm
yr^{-1}})(v_w/10^3~~{\rm km/s})$ and equivalent to $A_*=1$ in the
notation commonly used for GRBs \citep{clf03}.  On this same timescale
the radio emission from the SN, which is produced by a spherical shock
interacting with the circumstellar medium, will fade significantly
with a power-law decay index of $\alpha\approx 1.3-1.6$ \citep{lc99}.

Particle acceleration within the jet produces bright synchrotron emission
which causes a rapid rise of the radio luminosity as the jet comes into
our line-of-sight. 
Making the typical assumption that the energy
is partitioned between the electrons and the magnetic field, and that
these fractions ($\epsilon_e$ and $\epsilon_B$) are constant
throughout the evolution of the jet, the spectral luminosity is given
by \citet{wax03} as:

\begin{eqnarray}
L_{\nu} & = & 1.7\times 10^{30} (3\epsilon_e)(3\epsilon_B)^{3/4} \times \\
& &  \frac{(\dot{M}_{-5}/v_{w,3})^{9/4}}{E_{51}^{1/2}} \left( \frac{\nu}{10~{\rm GHz}} \right)^{-1/2} \left( \frac{t}{t_{oa}} \right)^{-3/2}~{\rm erg/s/Hz} \nonumber
\label{eqn:lum}
\end{eqnarray}

\noindent
On timescales $t \le t_{\rm oa}$ the spectral luminosity of the jet
rises steeply as $t^{+3.9}$ \citep{gpk+02} and during $t \ge t_{\rm
oa}$ it decays as $t^{-1.5}$, assuming an electron spectral index
$p=2$.  As indicated by Eqn 1, an off-axis GRB jet is
most easily detected at radio frequencies.

Late-time radio observations of SN\,1998bw can therefore be used to
impose constraints on the parameter space for an associated off-axis
GRB jet.  In modeling the evolution of the putative jet emission, we
consider four model parameters, $E_{51}$, $\dot{M}_{-5}/v_{w,3}$,
$\epsilon_e$ and $\epsilon_B$, for which we adopt values comparable
to those inferred for for cosmological GRBs.
The kinetic energy
of the jet and the energy fraction in electrons
are fairly well constrained by broadband observations
of GRB afterglows, so we assume typical values for these parameters of
$E_{51}=1$ and $\epsilon_e=0.1$ \citep{fks+01,pk01b,yhs+03}.  The
energy fraction in magnetic fields is significantly less constrained,
however, showing a typical range of values between $\epsilon_B =
0.002 - 0.25$ \citep{yhs+03}.  
This range imposes a significant uncertainty in the
luminosity of an off-axis jet, through Eqn 1.

The model is also strongly dependent on the mass loss rate of the
progenitor -- generally believed to be a massive Wolf-Rayet (WR) star.
Mass loss rates inferred from broadband modeling of GRBs are typically
$\dot{M}_{-5}/v_{w,3}\approx 0.1 - 1$ \citep{clf03,pk02,yhs+03}, yet
local WR stars have observed mass loss rates significantly larger, on the
order of $\dot{M}_{-5} \approx 0.6-9.5$
\citep{cgv03}.  The uncertainty associated with the progenitor mass loss
rate maps to a large range in the predicted peak time and 
luminosity of an off-axis jet.  Thus the uncertainty in
$\dot{M}_{-5}/v_{w,3}$ and $\epsilon_B$ produces a degeneracy in the
model which we are able to constrain using late-time radio observations
of local Ibc SNe.

\subsection{The $\epsilon_B - \dot{M}_{-5}/v_{w,3}$ Degeneracy} 
\label{sec:deg}

Figure~\ref{fig:lum_limits} shows the predicted radio luminosity
evolution of an off-axis GRB jet based on Eqn 1 and
using values of $\dot{M}_{-5}/v_{w,3}=0.01, 0.1, 1$ and $10$.  This
plot demonstrates how the uncertainty in $\epsilon_B$ produces 
a family of lightcurves for each $\dot{M}_{-5}/v_{w,3}$ value.

\begin{figure}
\plotone{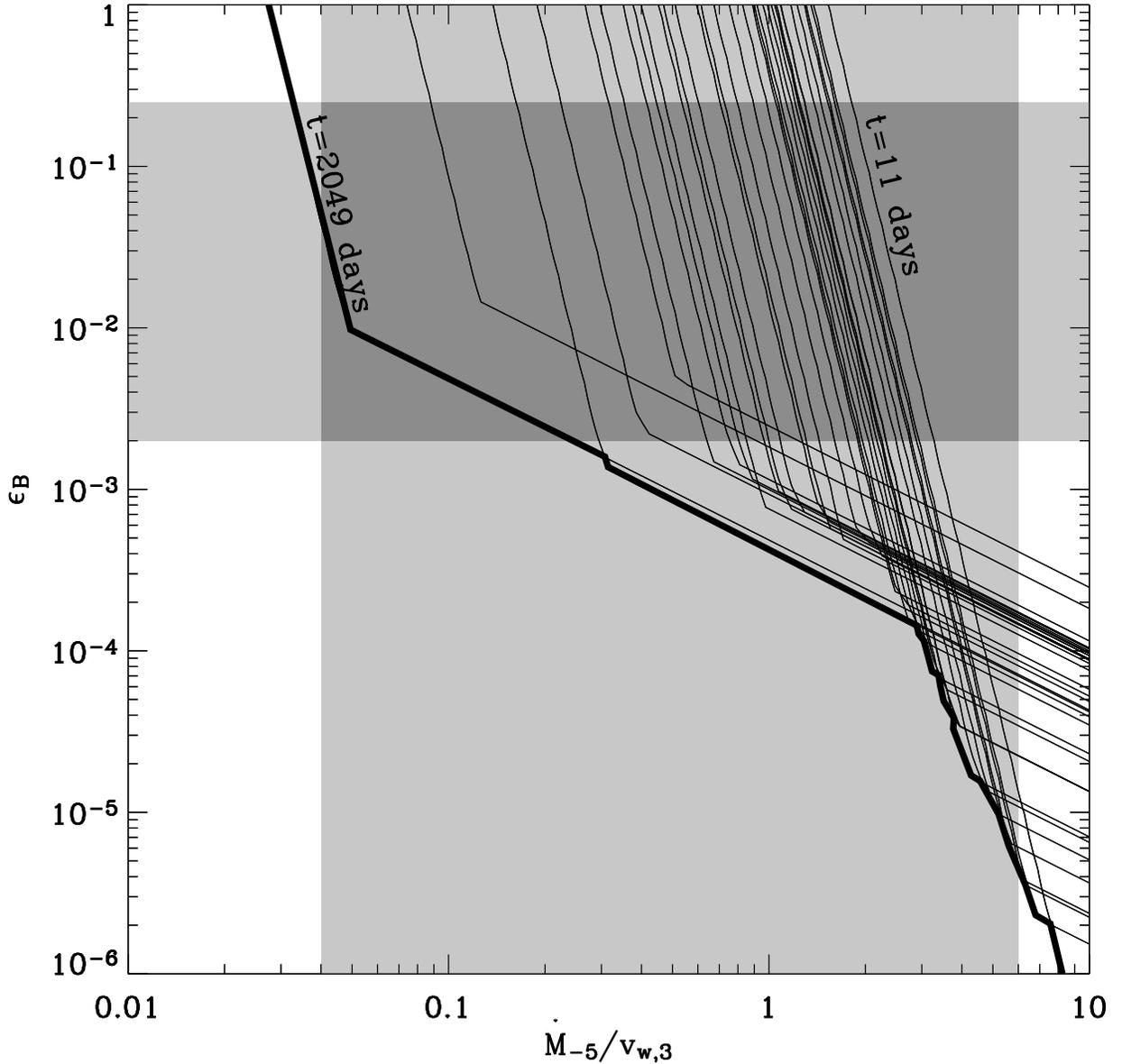}
\caption{Constraints on the off-axis GRB jet model for SN\,1998bw derived from
radio observations.  Every flux measurement allows us to
exclude the region of $\epsilon_B - \dot{M}$ space {\it rightward} of
the corresponding contour since this region produces a jet which is
{\it brighter} than the observed radio emission at that epoch.
The break in each contour marks the $\epsilon_B$ and
$\dot{M}_{-5}/v_{w,3}$ values that predict a peak jet luminosity equal to 
that observed.
Above the break, a contour constrains a jet with rising flux, while
below the break it constrains one with decaying emission. 
The sum of the excluded regions rules out the area rightward of the thick
solid line.  The region of reasonable parameter values for the
putative jet is bracketed by $\dot{M}_{-5}/v_{w,3}=0.04 - 6$ as
derived from modeling of SN\,1998bw radio light curves, and by
$\epsilon_B=0.002-0.25$ as inferred for cosmological GRBs.  Our
observations of SN\,1998bw allow us to rule out the majority of
this region.}
\label{fig:sn98bw}
\end{figure}

The full set of constraints is derived through
investigation of the two-dimensional parameter space of
$\epsilon_B - \dot{M}_{-5}/v_{w,3}$.  Figure~\ref{fig:sn98bw} shows
how the predicted spectral luminosity of the putative jet
at time, $t$, maps to a single contour in this parameter
space.  
By comparing the observed luminosity of SN\,1998bw at each epoch
with the jet luminosity predicted for that time, we
exclude the region of parameter space {\it rightward} of the
corresponding contour since this region
produces a jet which is {\it brighter} than
the observed SN\,1998bw emission at that epoch. 

The region of reasonable parameter space for the putative jet is
bracketed by $\dot{M}_{-5}/v_{w,3}=0.04 - 6$ as derived from modeling
of SN\,1998bw radio light curves
\citep{lc99,wax04}\footnote{\citet{lc99} prefer
$\dot{M}_{-5}/v_{w,3}=6$ for fitting the rise of the radio light
curves, while \citet{wax04} prefers $\dot{M}_{-5}/v_{w,3}=0.04$ for
fitting the X-ray light-curve decay.}, and by $\epsilon_B=0.002-0.25$
as inferred for most cosmological GRBs.  The summed constraints (from
$t=11$ to $2049$ days) rule out the majority of this region.

\section{Local Type Ibc Supernovae: Further Constraints}
\label{sec:constraints}

The search for an off-axis jet can similarly be carried out toward
other local ($d_L < 100$ Mpc) Type Ibc SNe for which there are
late-time ($t\sim 1-20$ year) radio observations.  Eight type Ibc SNe
were taken from \citet{bkf+03}: seven upper limits (SN\,2001B,
SN\,2001ci, SN\,2001ef, SN\,2001ej, SN\,2001is, SN\,2002J,
SN\,2002bm), and a detection (SN 2002ap) at $t\approx 0.5$ years.  We
have supplemented this sample with Very Large Array\footnote{The Very
Large Array is a facility of the National Science Foundation operated
under cooperative agreement by Associated Universities, Inc.}
archival observations of SN\,1983N, SN\,1984L, SN\,1985F, SN\,1990B,
SN\,1994I, and SN\,1997X taken at $t\approx 16.7,~2.9,~7.1,~0.8,~8.0$
and $2.4$ years, respectively. In addition, we include the recent
detection of SN\,2001em at $t\approx 2.4$ years \citep{svs+04} -- the
only SN within this sample for which there are no early time radio 
observations.  The SNe data are
plotted in Figure~\ref{fig:lum_limits}.  With the exception of
SN\,2001em, all of the late-time observations are significantly
fainter than the $\dot{M}_{-5}/v_{w,3}=1$ off-axis jet prediction and
nine SNe constrain $\dot{M}_{-5}/v_{w,3}\le 0.1$.

\begin{figure}
\plotone{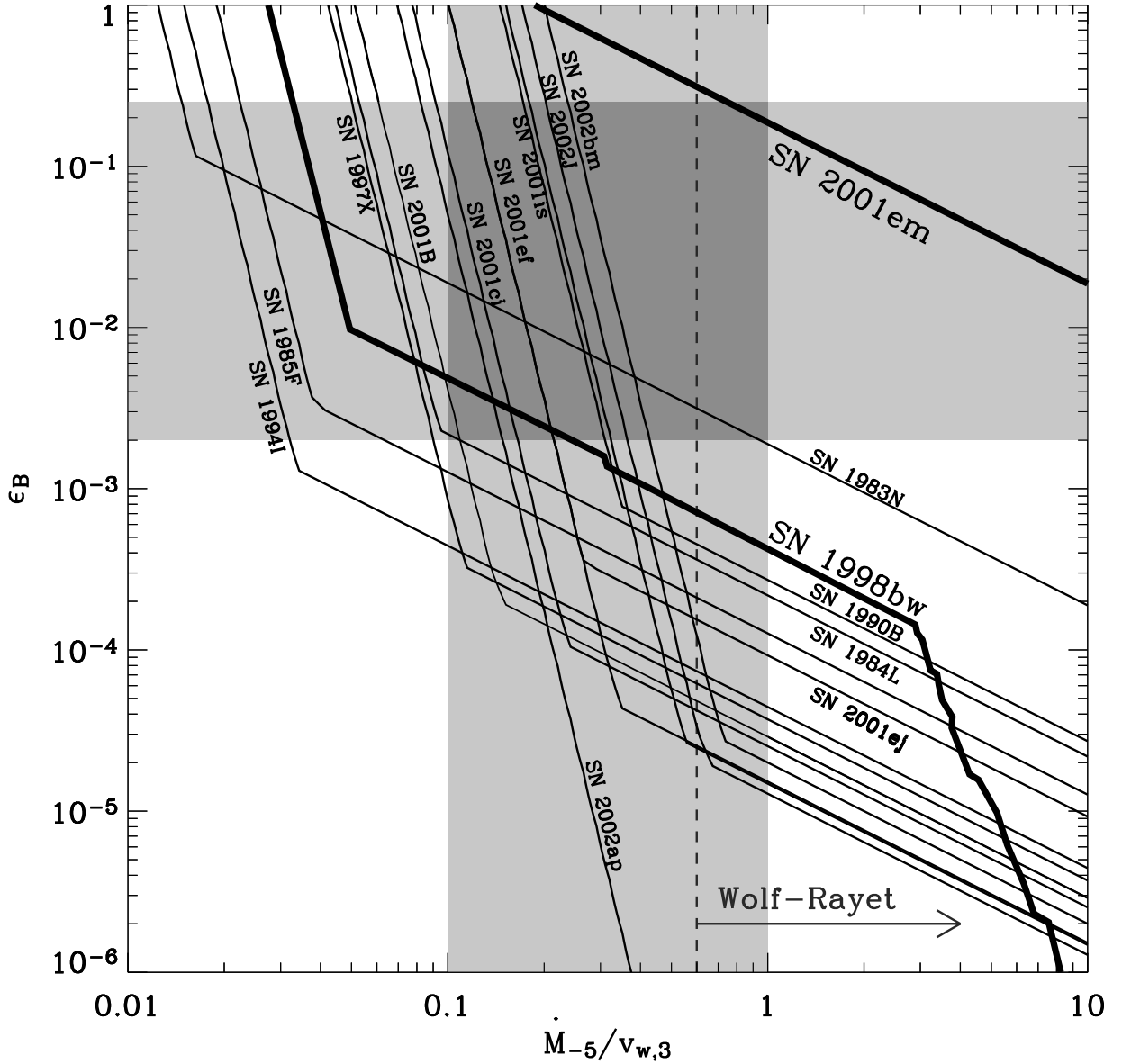}
\caption{Constraints on the off-axis jet model for GRBs derived from
late-time observations of local Ibc SNe.  Our observations allow us to
exclude the region of $\epsilon_B - \dot{M}$ space {\it rightward} of
each SN contour since this region produces a jet which is {\it
brighter} than the observed limit.  The region of reasonable parameter
values for an off-axis jet is roughly bracketed by
$\dot{M}_{-5}/v_{w,3}=A_*=0.1-1$ and $\epsilon_B=0.002-0.25$ as
typically inferred inferred from broadband modeling of cosmological
GRBs.  Four of the SNe contours rule out this entire shaded region
while eleven place significant constraints.  The recent detection of
SN\,2001em is consistent with a typical GRB jet only if the mass loss
rate is $\dot{M}_{-5}/v_{w,3}\approx 1$ (for $\epsilon_B\approx
0.2$).  The low end of the local distribution of observed Wolf-Rayet
mass loss rates is also shown. }
\label{fig:mdot_epsb}
\end{figure}

Using the same method applied in \S\ref{sec:sn98bw}, we
derive constraints for these 15 SNe.  The resulting contours are
plotted with SN\,1998bw in Figure~\ref{fig:mdot_epsb}.  For SNe with
later observations, the constraints on $\epsilon_B$ values improve, as
demonstrated by the contours for SN\,1994I and SN\,1983N.  The more
robust constraints are provided by the faintest
luminosity limits given by the nearest type Ic
supernovae, SN\,1994I, SN\,2002ap, SN\,1983N and SN\,1985F, all
at $\lesssim 8$ Mpc.  Additional constraints
are derived from the SNe for which we detect late-time
emission, SN\,1994I and SN\,2002ap\footnote{This rules out the Totani
(2003) model for SN2002ap predicting the
radio flux to increase rapidly at $t\approx 0.46$ years due to an
off-axis jet. However, this prediction assumes
$\dot{M}_{-5}/v_{w,3}\approx 0.3$ whereas lower values could delay the
onset of re-brightening.}, since they show evidence that the
radio luminosity is decaying with respect to earlier lightcurve
measurements (see \citealt{bkc02}) and are thus inconsistent with {\it
rising} off-axis jet emission.  

The region of reasonable parameter space for an off-axis jet is
roughly bracketed by $\dot{M}_{-5}/v_{w,3}=A_*=0.1-1$ and
$\epsilon_B=0.002-0.25$ as typically inferred from broadband modeling
of cosmological GRBs \citep{pk02,clf03,yhs+03}.
Figure~\ref{fig:mdot_epsb} shows that four of the SNe (1994I, 1985F,
1997X, 1990B) have contours that independently rule out this entire
parameter space while eleven place significant constraints.  
The late-time radio emission of SN\,2001em is consistent with
a typical GRB jet only if the mass loss rate is
$\dot{M}_{-5}/v_{w,3}\approx 1$ (for $\epsilon_B\approx 0.2$).
Continued radio monitoring of SN\,2001em will reveal the origin of
the emission, which may simply result from a late peaking
radio SN within a dense circumstellar medium \citep{wpv+02}.

\section{Discussion and Conclusions}

Late-time observations of SN\,1998bw ($t\simeq$5.6 yrs) have
allowed us to test the hypothesis that GRB\,980425 was a standard GRB
viewed far away from the jet axis. Our measured upper limits at 1384 and
2368 MHz are consistent with the continued power-law decay of the SN
emission.  These limits imply an off-axis jet is only
plausible if the normalized mass loss rate of the progenitor star is 
$\dot{M}_{-5}/v_{w,3} \le 0.04$ (for $\epsilon_B \ge 0.1$).  
This is $\sim 20-200$
times smaller than the observed mass loss rates for local Wolf-Rayet
stars \citep{cgv03} and is below the range typically observed in GRBs.
Larger mass loss
rates are possible but only if the energy fraction in magnetic fields
is low ({\em i.e.}, $\epsilon_B \lesssim 10^{-3}$).  

Even tighter constraints are derived for the off-axis jet model when we
examine a larger sample of local Ibc SNe. The low luminosity
limits derived for this sample require values of
$\dot{M}_{-5}/v_{w,3}\approx 0.01-0.1$ or  $\epsilon_B \lesssim  10^{-3}$
which are below values for typical GRBs.
The absence of any late-time radio emission can therefore be used to
put a limit on the fraction of core-collapse SNe that produce
collimated, relativistic outflows. Our results imply that
off-axis jets from nearby SNe are rare ($\lesssim $6\%)
with the possible exception that the radio emission from SN\,2001em is
due to a GRB jet.

This conclusion complements the findings of \citet{bkf+03} who
constrained the GRB/SN fraction through a radio survey of local 
Ibc SNe at early time.
\citet{bkf+03} used early, bright radio emission as a proxy for 
relativistic ejecta, as in the case for SN\,1998bw.
After studying 33 local SNe with detection limits $10^3$ times
fainter than SN\,1998bw, \citet{bkf+03} found no
evidence for relativistic ejecta in any of the SNe 
observed, thereby constraining the GRB/SN fraction to $\lesssim 3\%$.

Taken together, these results support a view that SN\,1998bw was a
rare and unusually energetic SN -- distinct from local SNe and
GRBs. In this scenario, the characteristics of SN\,1998bw/GRB\,980425 are not
dictated by the observer's viewing angle, but rather by the properties
of its central engine. SN\,1998bw was an engine-driven explosion
 \citep{lc99}, in which 99.5\% of the kinetic energy ($\sim 10^{50}$
erg) was coupled to mildly ($\Gamma\approx 2$) relativistic
ejecta \citep{kfw+98}, while a mere 0.5\% was detected in the
ultra-relativistic ($\Gamma\approx 100$) flow.  In contrast,
GRBs couple most of their energy to relativistic $\gamma$-rays.  
The observed diversity of cosmic explosions (SNe, X-ray
flashes, and GRBs) may therefore be explained with a standard energy
yield, but with a varying fraction of that energy given to
relativistic ejecta \citep{bkp+03}.

We thank Edo Berger, Sarah Yost and Eli Waxman for helpful
discussions.  AMS is supported by the NSFGRFP.


\clearpage

\clearpage

\clearpage

\end{document}